# New neutron detector based on Micromegas technology for ADS projects

Samuel Andriamonje[a*], Grégory Andriamonje[b], Stephan Aune[a], Gilles Ban[c], Stéphane Breaud[d], Christophe Blandin[d], Esther Ferrer[a], Benoit Geslot[d], Arnaud Giganon[a], Ioannis Giomataris[a], Christian Jammes[d], Yacine Kadi[e], Philippe Laborie[c], Jean François Lecolley[c], Julien Pancin[a], Marc Riallot[a], Roberto Rosa[f], Lucia Sarchiapone[e], Jean Claude Steckmeyer[c], Joel Tillier[c]

[a]*CEA-Saclay, DSM/DAPNIA, F- 91191 Gif-sur-Yvette , France*
[b]*IXL - Université Bordeaux 1 - BAT. A31 - 351 cours de la Libération - F-33405 Talence Cedex, France*
[c]*CNRS/IN2P3 LPC Caen, 6 Boulevard Maréchal Juin,F-14050 Caen Cedex, France,*
[d]*CEA/DEN/ Cadarache, 13108 Saint-Paul Lez Durance-France,*
[e]*CERN CH 1211 Geneva-Switzerland*
[f]*ENEA-Casaccia, Via Anguillarese, 00060 Roma, Italy*

**Abstract**

A new neutron detector based on Micromegas technology has been developed for the measurement of the simulated neutron spectrum in the ADS project. After the presentation of simulated neutron spectra obtained in the interaction of 140 MeV protons with the spallation target inside the TRIGA core, a full description of the new detector configuration is given. The advantage of this detector compared to conventional neutron flux detectors and the results obtained with the first prototype at the CELINA 14 MeV neutron source facility at CEA-Cadarache are presented. The future developments of operational Piccolo-Micromegas for fast neutron reactors are also described.

---

[*] Corresponding author. Tel.: +33-1-69-08-55-86; fax: +33-1-69-08-75-84; e-mail: sandriamonje@cea.fr (S. Andriamonje).

## 1. Introduction

In the ADS (Accelerator Driven System) project, one of the most important aspects needed for approval of the demonstrator is the experimental verification of the simulation. Of particular interest is the determination of the neutron spectrum (i.e. neutron flux as a function of the neutron energy) for different configurations of the sub-critical device. The neutron flux in ADS consists of neutrons produced via spallation reactions in the target and fissions from the multiplying blanket.

Unfortunately the neutron spectra cannot be measured using only one type of detector. To cover the complete energy range of the produced neutrons, a new neutron detector concept based on Micromegas technology has been developed.

The requirements for neutron spectrum measurements in the ADS project and the description of the new detector configuration dedicated to the neutron flux (i.e. neutron spectra) measurement in the present ADS project are given. In order to be placed inside an empty rod of a reactor this detector needs to be compact (3.5 cm x 3.5 cm x 3.5 cm) hence the origin of the Piccolo-Micromegas name.

The advantage of this detector compared to conventional neutron flux detectors and the results obtained with the first prototype at the CELINA 14 MeV neutron source facility at CEA-Cadarache are finally presented.

## 2. Simulated neutron spectra in ADS project

One example of the possible subcritical configuration based in TRIGA-ADS project [1], is shown in Fig. 1. The central rod and the rods named ring B are reserved for the spallation target and its cooling system. The measurement can be only achieved starting from ring C. The distribution of neutrons in three-dimensional space as a function of energy and time is simulated using the innovative simulation codes (FLUKA and EA-MC [2,3] at CERN, and MCNP-4C [4] at ENEA/Casaccia).

The analysis of neutron flux spectra is a necessary step in order to characterize the system. The TRIGA-ADS project is essentially a thermal system, cooled and moderated by water and using a fuel matrix rich in hydrogen.

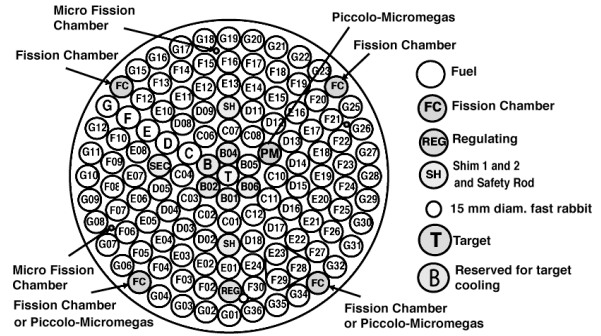

Fig. 1. Schematic view of the Fuel configuration for Sub-critical operation with possible locations of the neutron flux detectors.

Despite the high fuel enrichment (20 % $^{235}$U) and the rather small pitch between fuel elements (~ 2.35 mm for ~ 18 mm radius fuel elements), the abundance of this element guarantees the thermalization of the fission neutrons inside the fuel, therefore defining the very characteristic two-humped spectrum (the thermal and the high-energy humps). Moreover, the presence of a water buffer region (empty fuel channels in ring B) causes a prompt energy loss for the spallation neutrons and their diffusion, increasing the efficiency of the source neutrons.

Fig. 2 shows the neutron energy spectrum for different regions of the core for $k_{src}$ = 0.97. The target presents a harder neutron flux and the energy spectrum shows a high-energy peak around 2 – 3 MeV (spallation peak) and the resonances of tantalum (in particular, the ~15,000 barn resonance at 4 eV). The thermal hump (Maxwell distribution) is due to the neutrons reflected into the centre of the core. The fuel regions present similar neutron energy spectra, with decreasing integral value with the distance from the centre of the core.

The presence of the spallation source produces a relative energy shift between the high-energy flux (target region) to the thermal distribution (reflector region) permitting, to a limited extent, the analysis of different elements under different spectrum conditions.

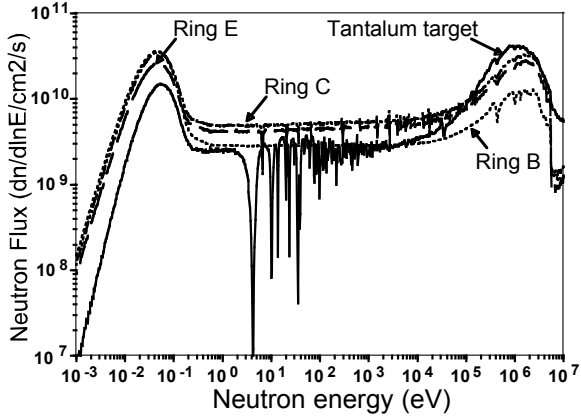

Fig. 2. Neutron flux spectra at selected locations of the TRIGA-ADS reference configuration.

## 3. Micromegas as neutron detector

### 3.1. Brief description of MICROMEGAS technology

The general description of the principle of Micromegas (MICRO-MEsh-GAseous Structure) technology can be found in references [5,6]. We describe here the typical aspect of the Piccolo-Micromegas detector for in-core measurement in nuclear reactor.

The principle of the Micromegas neutron detector used in n_TOF experiment [7,8] is reported in Fig. 3. As shown in this figure, the amplification occurs between the mesh plane and the microstrip plane.

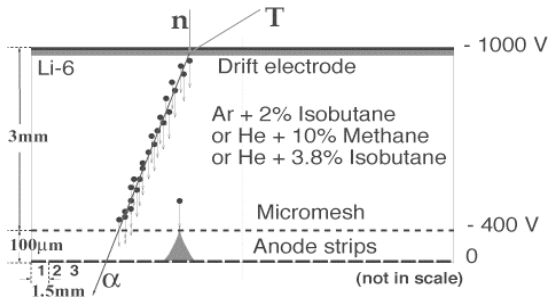

Fig. 3. The principle of Micromegas for neutron detector used in n_TOF experiment [8].

The device operates as a two-stage parallel plate avalanche chamber. Ionization electrons, created by the energy deposition of an incident charged particle in the conversion gap, drift and can be transferred through the cathode micromesh, they are then amplified in the small gap, between anode and cathode, under the action of the electric field which is high in this region. The electron cloud is finally collected by the anode microstrip, while the positive ions are drifting in the opposite direction and are collected on the micromesh. For this experiment, appropriate neutron/charged particle converters have been used (see Fig. 3): (i) $^6$Li(n,α) for the neutron having an energy up to 1 MeV, (ii) H(n,n')H and $^4$He(n,n')$^4$He for higher energy neutron.

This method has been extended to the measurement of neutron flux in-core environment. One of the main qualities of the detector is its high resistance to the radiation. Several experiments have shown that the Micromegas detector meets with this requirement. Only 10% variation of the gain has been observed after an irradiation of the detector of 24 mC/mm$^2$, this is equivalent of ten years irradiation of the detector on LHC (Large Hadron Collider) project [9].

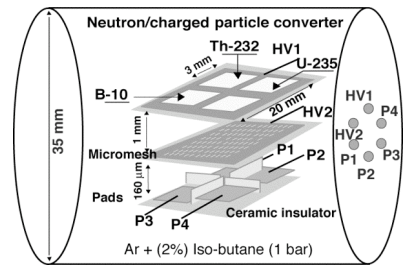

Fig. 4. Schematic view of the Piccolo-Micromegas detector (in horizontal position) for neutron flux measurements.

### 3.2. Description of Piccolo-Micromegas

The principle of Piccolo-Micromegas detector is shown in Fig. 4. The different materials, *stainless steel* and ceramic, have been chosen in order to respect the following conditions:
- low activation with high neutron flux,
- small perturbation of the neutron flux,
- homogeneity of the dilatation coefficients,
- low out gassing for a sealed gaseous detector,

**The chamber body of the detector** is a *stainless steel* cylinder with 35 mm of external diameter, 30 mm of internal diameter and 35 mm of height and closed in the bottom part. **The cover** is made of *stainless steel* equipped by a special piece for handling the detector inside the reactor rod and six special connectors stuck or welded by a special technique already used for the micro fission chamber [10]. **The drift electrode** is composed of three neutron/charged particle converters: $^{10}B$, $^{235}U$, $^{232}Th$ and one empty for the recoil nucleus of H of Ar-$C_4H_{10}$ filling gas. According to the neutron interaction cross section: - the $^{235}U$ is used for monitoring, since the fission cross section of $^{235}U$ from thermal to several MeV neutron energy is well known, - the $^{10}B$ for the measurement of epithermal neutron flux (up to several keV) - the $^{232}Th$ for the measurement of fast neutrons having an energy larger than the fission threshold (~ 1 MeV) and the H+ recoil for the neutron having an energy greater than the energy threshold of the detector (several keV). **The Micromesh** is a *stainless steel* woven with 19 µm wires. 500 lines per inch pitch insure the transparency. The drift gap (1 mm) and the amplification gap (160 µm the distance between the Micromesh and the **anode pad**) are ensured by an insulating part in ceramic opening by four holes having the same diameter that the deposit sample.

## 4. Experiment test and first results

### 4.1. Characteristics of Piccolo-Micromegas detector

To determine the main characteristics of the new Micromegas detector, $^{55}Fe$ X-ray source and two types of gas mixture (Ar-10%$CO_2$ and Ar-2%$C_4H_{10}$) have been used. The pressure of the sealed detector has been set to one bar. The ionization power of fission fragment compared to recoil nucleus or alpha particle are very different. In order to adapt the dynamic range of the collected charge, a new method has been used. The amplification gap has been taken identical for the four pads (160 µm). The drift electrode is grounded. The mesh and the four reading pads have been polarized positively and individually. The signal from each pad is collected and shaped by a homemade fast amplifier. An oscilloscope-PC has been used for the data acquisition. The energy resolution of the detector at 5.9 keV $^{55}Fe$ X-ray is about 20%. The gains of the detector as a function of the different voltage applied on the different pads have been measured.

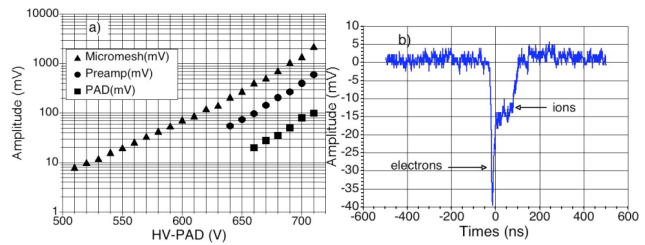

Fig. 5. a) Gain of the Piccolo-Micromegas detector versus HV-PAD for HV-Mesh = +110 V and the drift cathode in the ground and 1 bar Ar+$CO_2$, b) Example of the fission fragment pulse.

The results show clearly (Fig. 5a) that a very low polarization of the pad is largely sufficient for the observation of the signal from fission fragments (several MeV compared to 5.9 keV energy deposited by the X ray of $^{55}Fe$ radioactive source).

### 4.2. Experiments and results with neutron source

The first test of the Piccolo-detector has been performed at the 14 MeV neutron source CELINA facility at CEA-Cadarache. Two types of gas mixtures (cf. Paragraph 4.1) have been used to fill at 1 bar the sealed detector.

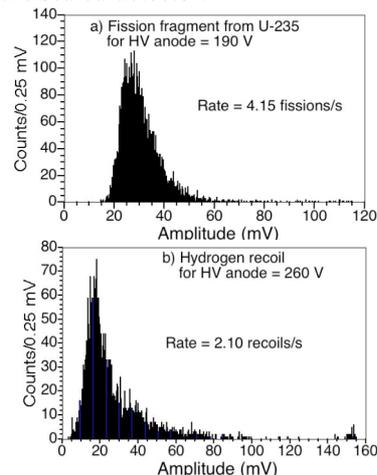

Fig. 6. a) Amplitude histogram of $^{235}U$ fission fragments, b) Amplitude histogram of H+ recoils.

An example of the pulse obtained from the fission fragment emitted by the interaction of a neutron with $^{235}$U is shown in Fig. 5b. The contribution of the fast electron and the slow ions are shown clearly in the figure. A special routine has been developed with Matlab [11] to analysis and sort the raw data for eventual spark and saturation.

Two histograms as a function of the amplitude of the pulse (proportional to the energy deposited by the charged particle) are reported in Fig. 6a and Fig. 6b.

The first one corresponds to the $^{235}$U fission fragments and the second one to the recoils of hydrogen. The gain used for the fission fragments is ~20 times lower than the one used for the H$^+$ recoils.

The resolution of the detector is not good enough to observe the two bumps of the fission fragments distribution.

*4.3. Comparison with simulation*

A full simulation of the Piccolo-Micromegas detector placed inside the CELINA-Cadarache neutron source facility has been performed using FLUKA [2]. The study has been centered on the particular case of $^{235}$U, because of the well-known fission cross section. The model used consisted of a 14 MeV neutron source moderated by graphite with an intensity of about $2 \times 10^9$ n/s/4π and samples of $^{235}$U and $^{232}$Th 50 μg each.

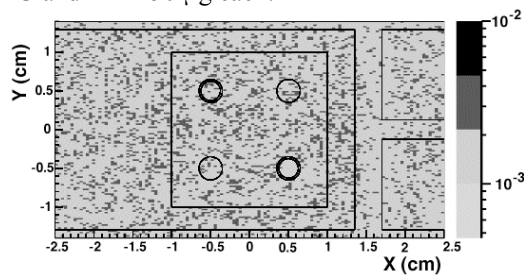

Fig. 7. Neutron flux distribution (arbitrary units) inside the Piccolo-Micromegas detector

The neutron flux distribution inside the Piccolo-Micromegas detector is shown in Fig. 7. The detector seems homogeneously illuminated by the 14 MeV neutron source placed a few tens of cm away, inducing some ~0.08 fissions/s per μg of $^{235}$U. This is in good agreement with the measured integral of the fission fragment spectrum (Fig. 6a).

## 5. Conclusion

A new neutron detector named Piccolo-Micromegas has been developed. The first results presented in this paper show clearly the performance and the ability of this detector to measure simultaneously from thermal to high-energy neutron flux in the nuclear reactor especially in ADS project.

The next step is the use of a special material for the connectors and the cable for operation at high temperature up to 300 °C.